\documentclass{icrc29}

\usepackage{graphicx,amssymb,amsmath,times}
\begin{document}

\title[Alternative energy estimation from the shower lateral
  distribution function]{Alternative energy estimation from the shower
  lateral distribution function} 

\author[V. de Souza, C. Escobar, J. Brito, C. Dobrigkeit and
  G. Medina-Tanco] {Vitor de Souza$^a$, Carlos Escobar$^b$, Joel Brito$^b$,
  Carola Dobrigkeit$^b$ and Gustavo
  Medina-Tanco$^a$ \\ 
{\it [a] Instituto de Astronomia e Geof\'{\i}sica, Universidade de S\~ao Paulo, Brasil} \\
{\it [b] Departamento de Raios C\'osmicos, IFGW, Universidade Estadual de Campinas, Brasil}
} 

\presenter{Presenter: C. Escobar (escobar@ifi.unicamp.br) bra-escobar-CO-abs2-he14-poster}

\maketitle

    
\begin{abstract}

The surface detector technique has been successfully used to detect
cosmic ray showers for several decades. Scintillators or Cerenkov
water tanks can be used to measure the number of particles and/or the
energy density at a given depth in the atmosphere and reconstruct the
primary particle properties. It has been shown that the experiment
configuration and the resolution in reconstructing the core position
determine a distance to the shower axis at which the lateral
distribution function (LDF) of particles shows the least variation
with respect to different primary particles type, simulation models
and specific shapes of the LDF. Therefore, the signal at this distance
(600 m for Haverah Park and 1000 m for Auger Observatory) has shown to
be a good estimator of the shower energy. Revisiting the above
technique, we show that a range of distances to the shower axis,
instead of one single point, can be used as estimator of the shower
energy. A comparison is done for the Auger Observatory configuration
and the new estimator proposed here is shown to be a good and robust
alternative to the standard single point procedure.

\end{abstract}

\section{Introduction}

Surface array detectors have been used to detect cosmic ray showers due
to a large number of properties, including their stability, large
detection areas and duty cycle. Nowadays, the arrival time of the particles in
the shower can be measured with a good resolution (tenths of ns) leading to an
excellent reconstruction of the primary particle direction ($>
1^\circ$) and therefore good precision in anisotropy studies. Besides
that,  ground array detectors have a well defined aperture resulting
in a straight forward determination of the cosmic ray spectrum.

However, since ground array experiments detect a sample of the shower
development at a single fixed depth, the energy reconstruction can not
be derived in a direct or calorimetric way. In fact, the most
important information from which the parameters of the primary
particle should be reconstructed is the lateral distribution of
particles in the shower. 
In 1969, Hillas et al. \cite{bib:hillas:s500} showed that at a given
distance of the shower axis, the fluctuations of the lateral
distribution function (LDF) due to intrinsic shower fluctuations are
minimized. In subsequent works \cite{bib:hillas:s500:2}, it was also shown that the measured
signal fluctuation is very small at a certain distance in despite of
the type of primary 
particles, Monte Carlo simulation models and specific LDF functions.   

The distance at which the fluctuations reach their minimum is a
convolution of the 
intrinsic shower-to-shower fluctuation, the experimental uncertainties
and the reconstruction procedures. One very important experimental
input is the array spacing which mathematically determines the
properties of the LDF fit \cite{bib:gustavo:ldf}.
The Haverah Park experiment \cite{bib:haverah}  used water Cerenkov tanks and determined
the distance of least fluctuation to be 600 meters. Operating with the
same ground array technique, the 
Pierre Auger Observatory \cite{Auger} has determined the distance of least
fluctuation to be 1000 meters.  

The scope of this work is to explore the possibility to use the
integral of the LDF within a given interval as an energy estimator. A
similar procedure is used by the KASCADE Collaboration
\cite{bib:kascade:ldf} in order to 
reconstruct the energy of the primary particle. We investigate
the fluctuation and the resolution of this parameter based on
simulation studies for showers with
energies from $10^{18}$ to $10^{19.5}$ eV.  
 
\section{Simulation and Reconstruction}
\label{sec:sim:rec}

We have used a simplified numerical simulation of the signal in a
water Cerenkov tank given by a parametrization of the lateral
distribution of the particles as a function of the primary energy as
explained in reference \cite{bib:gustavo:ldf}. Poissonian noise is
considered. We simulated a 1500 stations array with 1.5 km
spacing. Only stations with signal above 3.2 VEM which did not saturate
were considered in our analysis.

The signal and position of each station was used to fit three
different LDF types as given below: 

\vspace{0.5cm}
\underline{NKG-Type:}
\begin{equation}
S(r) = S(1000) \left( \frac{r}{1000} \right)^{-\beta - 0.2} \left(
\frac{r+r_s}{1000 + r_s} \right) ^{-\beta}
\end{equation}

\vspace{0.5cm}
\underline{Power Law (PL):}
\begin{equation}
S(r) = S(1000) \left( \frac{r}{1000} \right)^{-\nu}
\end{equation}

\vspace{0.5cm}
\underline{Monte Carlo inspired (MC):}
\begin{equation}
S(r) = 10^{A + Bx+Cx^2}
\end{equation}

\vspace{0.5cm}
where $\nu = 5.1 - 1.4 \times \sec \theta$, $\beta = 3.3 - 0.9 \times \sec \theta$ , 
$r_s = 700 m$ and $x = \log (r/1000)$.

\begin{figure}[t]
\begin{minipage}[t]{7.5cm}
\includegraphics*[angle=-90,width=1.0\textwidth]{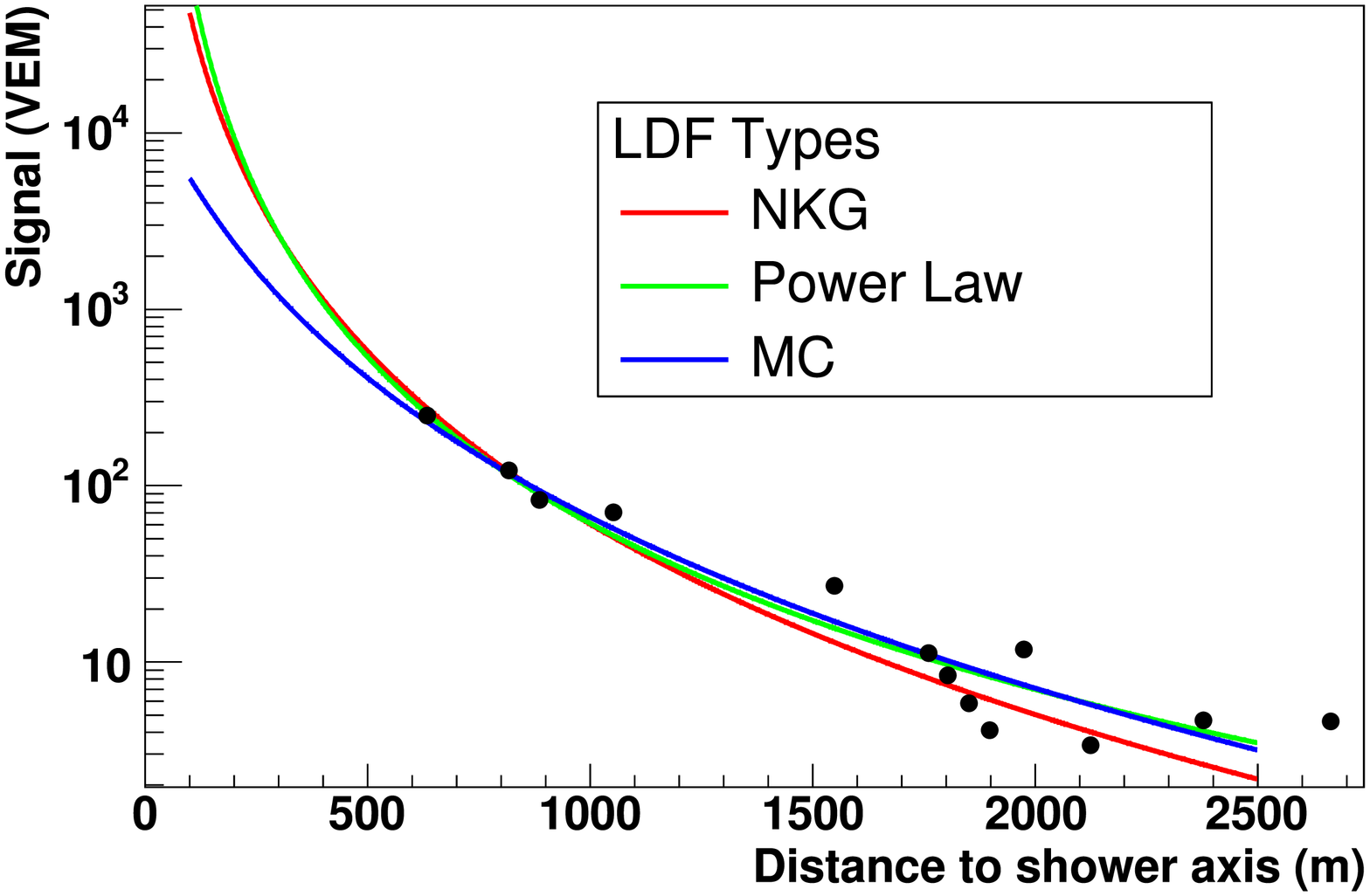}
\caption{\label{fig:ldf:ex} Example of a simulated event with energy
  $10^{19.5}$ eV and zenith angle $45^\circ$ and three fitted lateral functions.} 
\end{minipage}
\hfill
\begin{minipage}[t]{7.5cm}
\includegraphics*[angle=-90,width=1.0\textwidth]{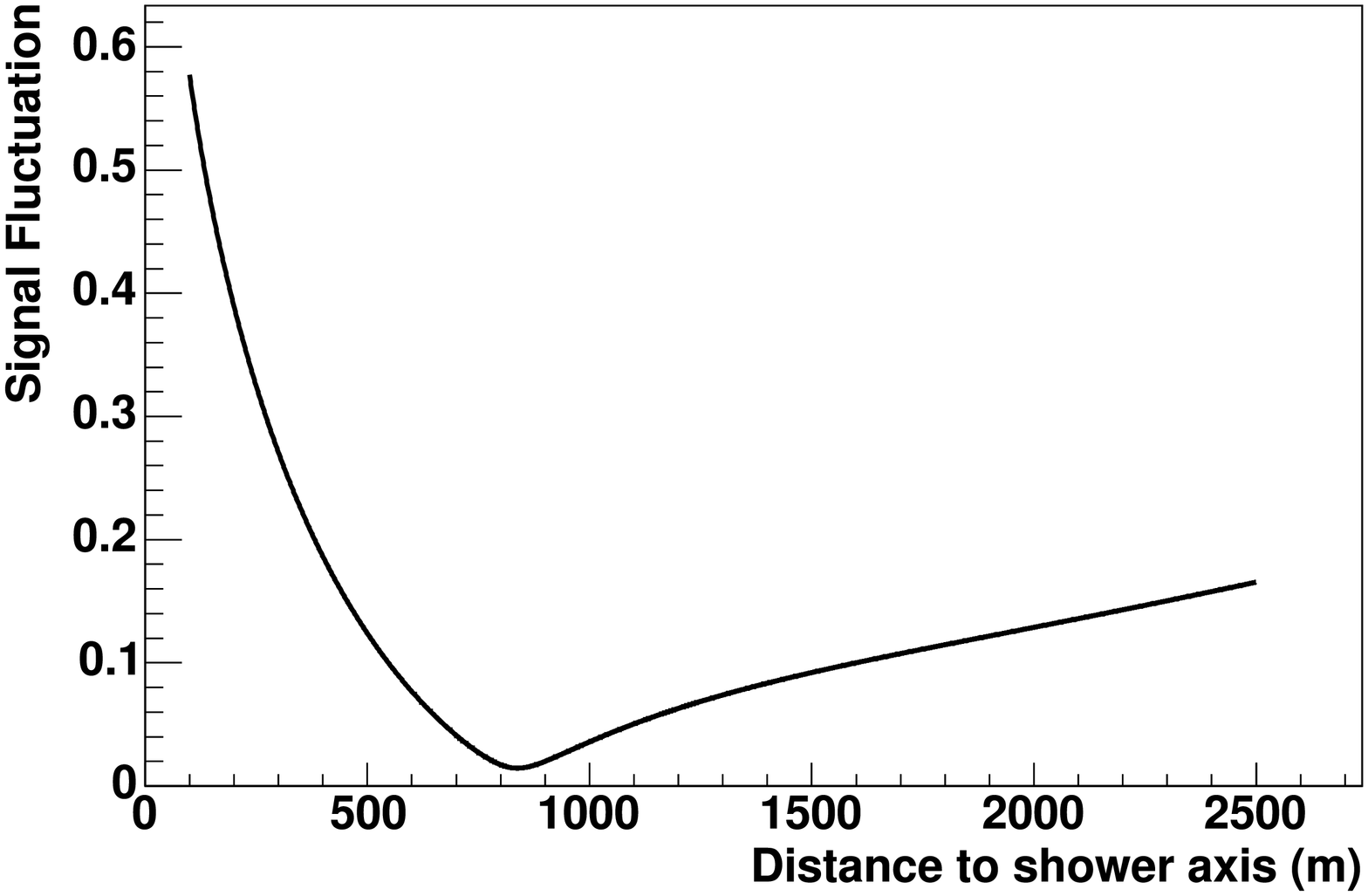}
\caption{\label{fig:desvio:ex} Fluctuations of the signal as a
    function of the distance to the shower axis. There is one clear
    minimum in the fluctuations around 800 m.}
\end{minipage}
\hfill
\end{figure}

These functions where suggested by the Auger Collaboration in reference
\cite{bib:icrc:roth}. Figure \ref{fig:ldf:ex} illustrates the procedure,
black dots correspond to the simulated signal in the stations which were
used to fit the three lines corresponding to each LDF type. Figure
\ref{fig:desvio:ex} shows the corresponding fluctuation where one
clear minimum around 800 m can be seen.

\begin{figure}[t]
\begin{minipage}[t]{7.5cm}
\includegraphics*[angle=-90,width=1.0\textwidth]{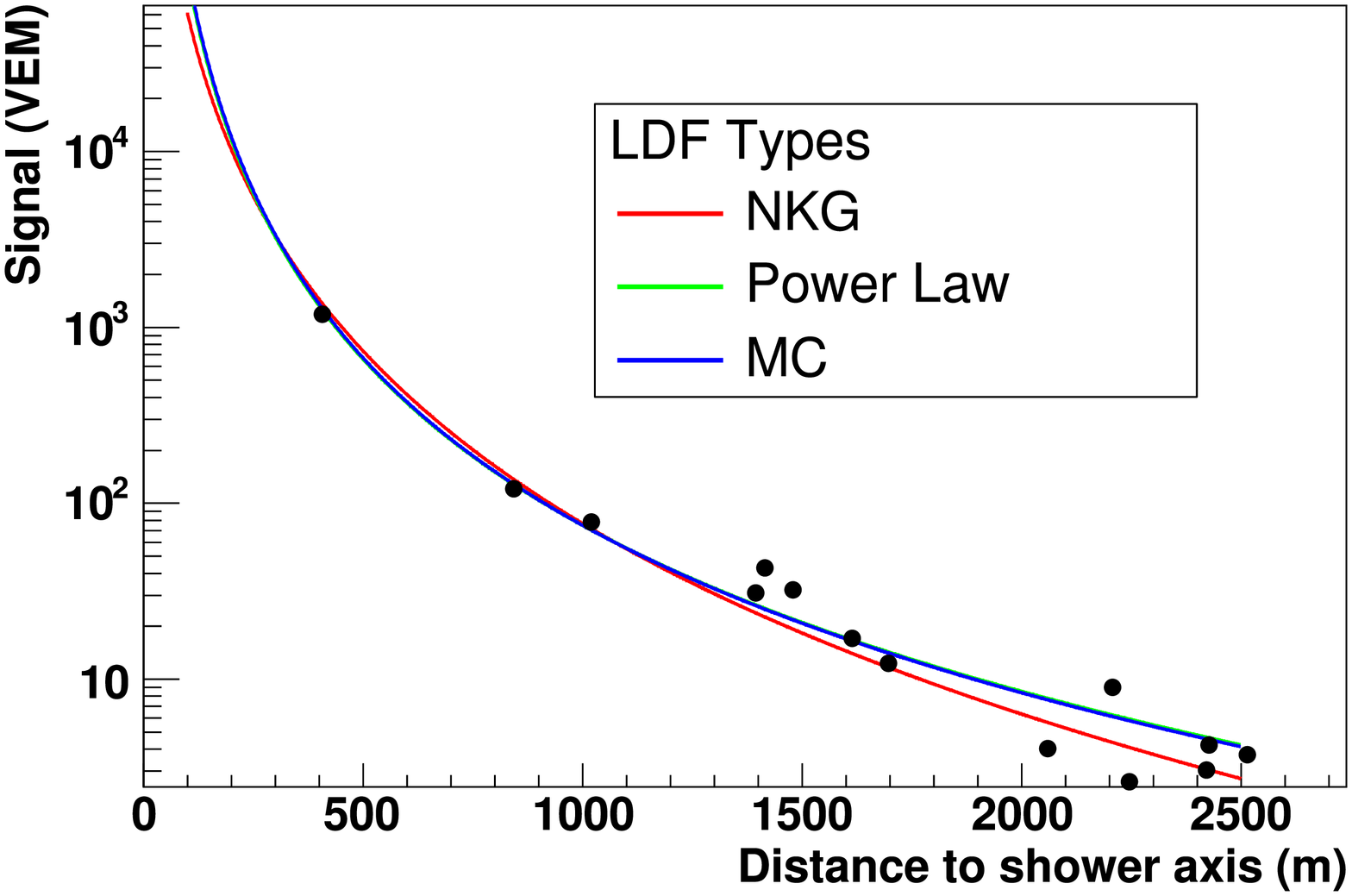}
\caption{\label{fig:ldf:ex2} Example of a simulated event with energy
  $10^{19.5}$ eV and zenith angle $45^\circ$ and three fitted lateral functions.} 
\end{minipage}
\hfill
\begin{minipage}[t]{7.5cm}
\includegraphics*[angle=-90,width=1.0\textwidth]{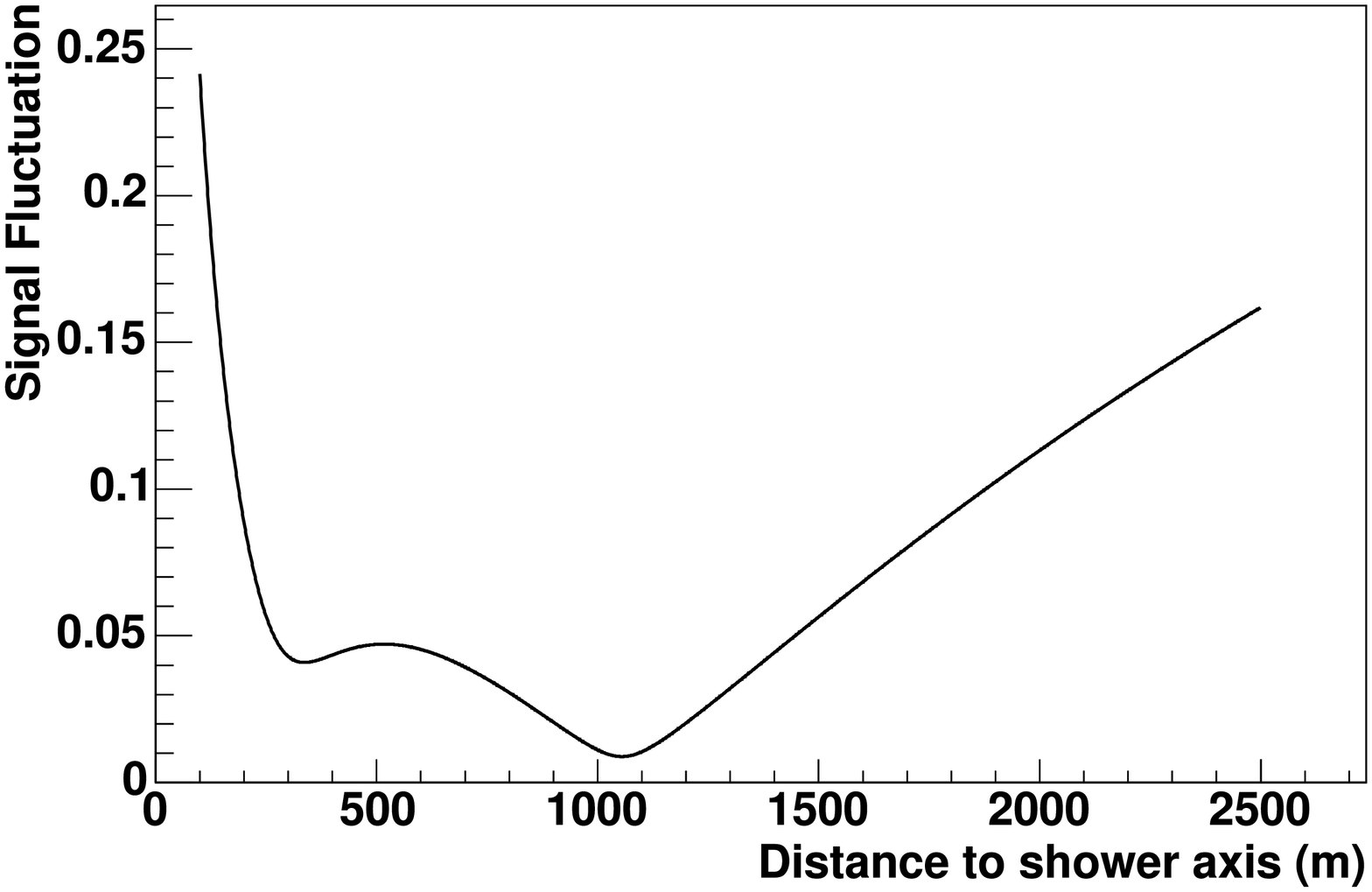}
\caption{\label{fig:desvio:ex2} Fluctuations of the signal as a
    function of the distance to the shower axis. Note the two minima
    in the fluctuation.}
\end{minipage}
\hfill
\end{figure}

It is also noticeable that the fluctuation is below 10\% for a large
interval of distances to the shower axis ($\sim$ from 500 to 1400 m). Figures \ref{fig:ldf:ex2} and
\ref{fig:desvio:ex2} show another example of LDF fit and the
corresponding fluctuation. In this case, the fluctuations
have two clear minima and an even larger interval of distances
($\sim$ from 200 to 1500 m) with fluctuations below 10\% is seen.

Following this procedure we simulated 100 proton showers corresponding to each
primary energy of $10^{18.5}$, $10^{19.0}$ and $10^{19.5}$ eV,
zenith angle of $45^\circ$ and
calculated for each event the distance of minimum fluctuation
($R_{opt}$).

\section{Results}

\begin{figure}[t]
\begin{minipage}[t]{7.5cm}
\vspace{-5.5cm}
\includegraphics*[angle=-90,width=1.0\textwidth]{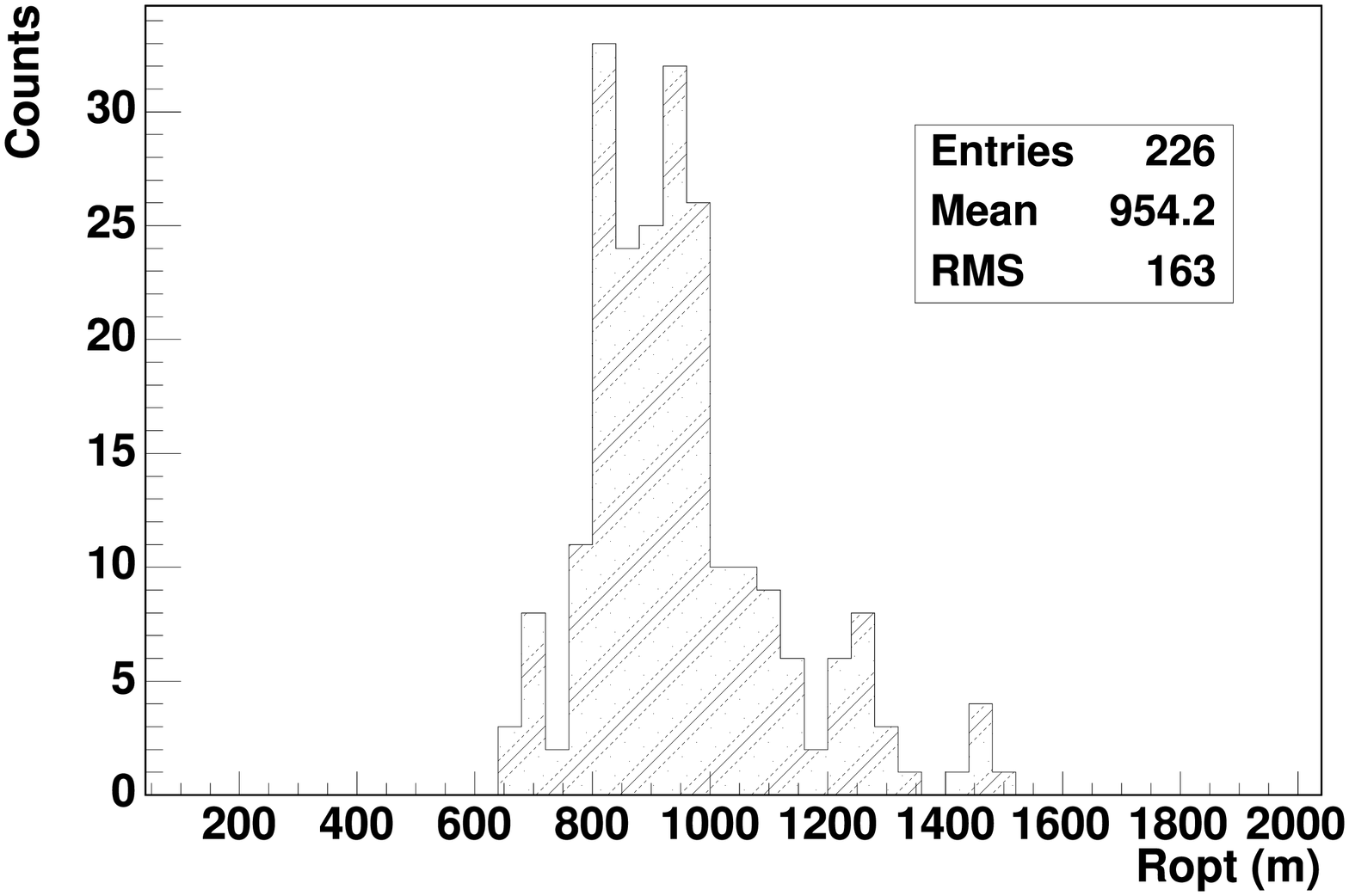}
\caption{\label{fig:ropt} Distribution of distance where the minimum
  fluctuations occur.}
\end{minipage}
\hfill
\begin{minipage}[t]{7.5cm}
\includegraphics*[width=1.0\textwidth]{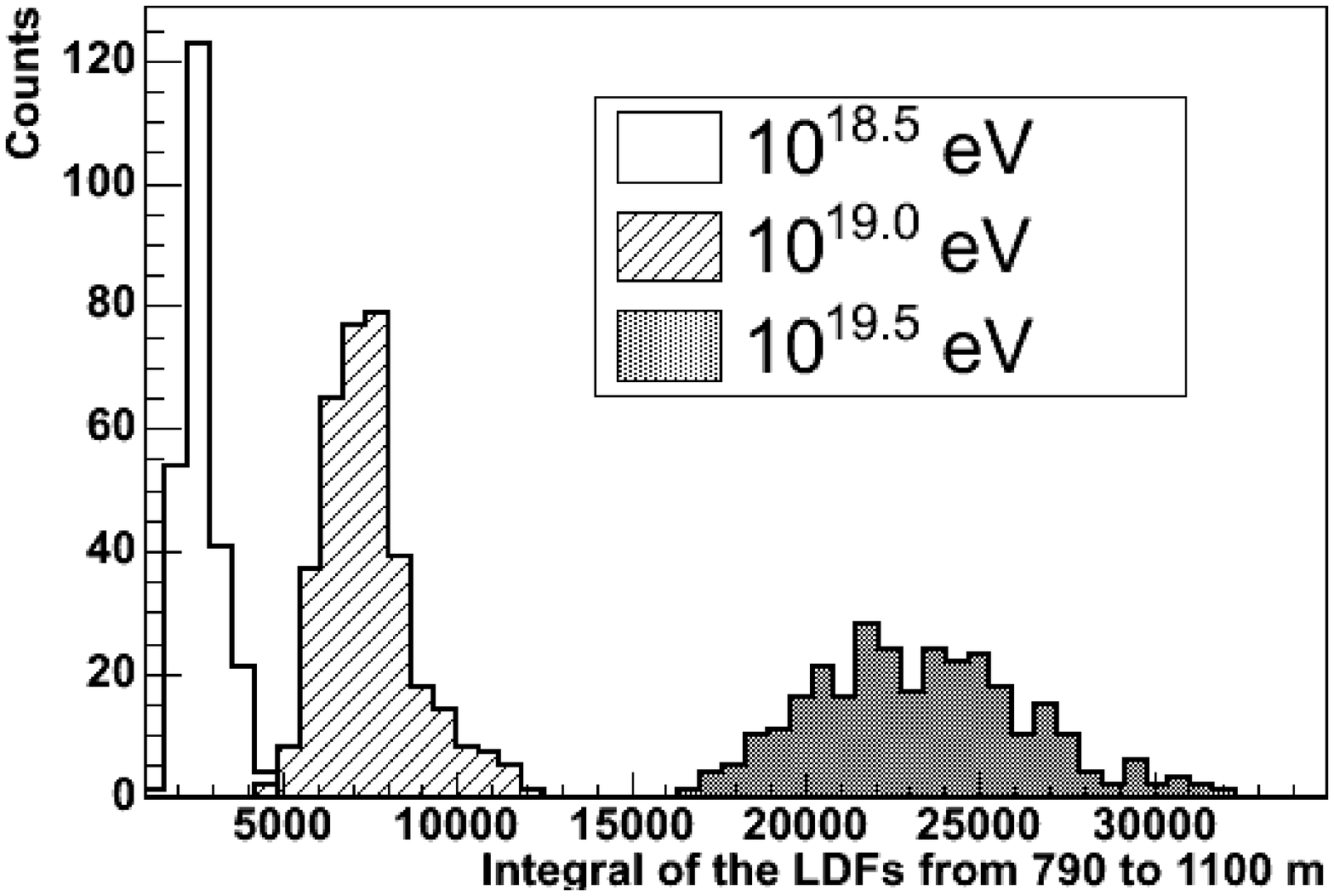}
\caption{\label{fig:integral} Distribution of the integral of the LDFs
from 790 to 1100 m.}
\end{minipage}
\hfill
\end{figure}

Figure \ref{fig:ropt} shows the distribution of $R_{opt}$ for all energies. If the
fluctuations as a function of distance showed two minima, only the absolute
minimum was included in the distribution shown in figure
\ref{fig:ropt}. We have only included showers for which we manage to fit all
three functions and events which had no saturated tanks. 

The use of one single point calculated to each event is under study.
However, the discussion in section \ref{sec:sim:rec} and figure
\ref{fig:ropt} suggest that the interval from $\bar R_{opt} - \sigma$ to
$\bar R_{opt} + \sigma$ could be used as an energy estimator because
the signal fluctuations are very small along this entire range.

We have also studied the  distributions of  $R_{opt}$ for each energy
$10^{18.5}$ , $10^{19.0}$ and $10^{19.5}$ eV
and we have verified the increase of  $\bar R_{opt}$, from
895 m, to  913 m and 1040 m, respectively. Indeed  figure \ref{fig:ropt}
shows two peaks, the first corresponding to the distribution of showers
with energies $10^{18.5}$ and $10^{19.0}$ eV and the second corresponding
to showers with energy $10^{19.5}$ eV.

Figure \ref{fig:integral} shows the value of the integral of the three
LDFs from 790 to 1110 m for the energies considered in this
analysis. The distributions of the signal at 1000 m (S(1000)) away from the
shower axis were also studied. Table \ref{tab:resultado} shows the
comparison of the spread of the distribution of the integral and the
S(1000).

\begin{table}[t]
\begin{center}
\begin{tabular}{|c|c|c|c|c|c|c|} \hline
     &  \multicolumn{2}{|c|}{$10^{18.5}$ eV} &
           \multicolumn{2}{|c|}{$10^{19.0}$ eV} &  \multicolumn{2}{|c|}{$10^{19.5}$ eV} \\ \hline
     &     Int  & S(1000)   &     Int  & S(1000)   &     Int  &
           S(1000)   \\ \hline
Mean   & 8336 &  7.30 &  23470 & 19.53 &  73670 &  60.94 \\ \hline
RMS    & 1891 &  1.57 & 4540 &3.47 &  9945 & 7.66 \\ \hline
Spread & 22\% &  21\%  & 19\% & 17\% & 13\% & 12\% \\ \hline
\end{tabular}
\caption{Mean, RMS and Spread (RMS over Mean) as a function of energy
  for the S(1000) and integral distributions.}
\label{tab:resultado}
\end{center}
\end{table}

\section{Conclusion}

Table \ref{tab:resultado} shows that the integral has a similar
spread as that of the S(1000) values. A more detailed work regarding hadronic
interaction models, several zenith angles and different primary
particles is in preparation and the integral procedure has shown to be
a promising good energy estimator. 

\section{Acknowledgments}

This paper was partially supported by the Brazilian Agencies CNPq and
FAPESP.


\begin{thebibliography}{99}

\bibitem{bib:hillas:s500}
M. Hillas et al., $11^{th}$ ICRC (1969) EAS-35

\bibitem{bib:hillas:s500:2}
M. Hillas et al., $12^{th}$ ICRC (1969) EAS-18

\bibitem{bib:gustavo:ldf}
G. Medina-Tanco, in these proceedings.

\bibitem{bib:haverah}
R. Coy, Astroparticle Physics 6 (1997) 263

\bibitem{Auger}
The Auger Collaboration, Nucl. Instr. Meth. A 523, 50 (2004).

\bibitem{bib:kascade:ldf}
The KASCADE Collaboration, Astroparticle Physics, 16 (2002) 245

\bibitem{bib:icrc:roth}

Markus Roth for the Pierre Auger Collaboration, $28^{th}$ ICRC pag. 333

\end{thebibliography}
\end{document}